\documentclass[pra,aps,preprint]{revtex4}
\usepackage{amssymb}
\usepackage{epsfig}
\usepackage{graphicx}
\begin{document}
\title{The Casimir force between dissimilar mirrors and the role of the surface plasmons}
\author{A. Lambrecht$^1$}
\author{I. G. Pirozhenko$^{2}$}
\affiliation{$^1$Laboratoire Kastler Brossel, CNRS, ENS,
UPMC - Campus Jussieu case 74, 75252 Paris, France\\
$^2$Bogoliubov Laboratory of Theoretical Physics, JINR, 141980
Dubna, Russia}
\begin{abstract}
We investigate the  Casimir force between two dissimilar plane
mirrors the material properties of which are described by Drude or
Lorentz models. We calculate analytically the short and long
distance asymptote of the force and relate its behavior to the
influence of interacting surface plasmons. In particular we discuss
conditions under which Casimir repulsion could be achieved.
\end{abstract}
\date{\today}
\maketitle

\section{Introduction}
The availability of modern experimental set-ups that allow accurate
measurements of surface forces between macroscopic objects at
submicron separations has stimulated increasing interest in the
Casimir effect \cite{Casimir:1948}. In his seminal paper Hendrik
Casimir calculated the force between two plane parallel perfectly
reflecting mirrors in vacuum and found the following expressions for
the Casimir force and energy per unit area
\begin{equation}
F_{\rm C}= -\frac{\hbar c\pi ^2}{240 L^4} \quad E_{\rm C}=
-\frac{\hbar c\pi ^2}{720 L^3}\label{Casimir}
\end{equation}
where sign conventions are such that a negative value of $F_{\rm C}$
and the energy correspond to attraction between the two plates. $L$
is the mirrors' separation.

The Casimir force between two dielectric plates was first derived
in~\cite{Lifshitz:Eng:1956}. In the original paper is was mentioned
that the spectrum of the problem includes both propagative and
evanescent waves. Later, a number of
authors~\cite{VanKampen:Eng:1968,Gerlach:1971,0038-5670-18-5-R01}
succeeded in obtaining the same result through the summation of the
modes corresponding to the solutions of the Maxwell equations which
decay exponentially in the direction normal to the plates. As a
consequence the entire force was interpreted as the interaction of
the surface plasmons. However, Schram~\cite{Schram:1973} noticed
that the surface modes alone do not yield the right expression. At
present we understand that the transition from mode summation to
contour integration made
in~\cite{VanKampen:Eng:1968,Gerlach:1971,0038-5670-18-5-R01,Schram:1973},
is the most delicate point in the calculation and a possible source
of this confusion. The modern point of view is that the surface
plasmon interaction dominates at short
separations~\cite{Barton:1979,Genet:2004,HenGreffet:2004}. For
plates described by a plasma model or plasma sheets it was shown
that the value of the Casimir force is the result of large
cancelations between plasmon and photon contributions at all
distances\cite{IntraLam:2005,Bordag:2005}. The interaction of the
surface modes between a sphere and a plate taking into account the
material properties was considered
in~\cite{Noguez:2004,Esquivel:2004}.

In the present paper we study the Casimir force between
\textit{dissimilar} plates with dispersion and assess the role of
 surface plasmons in this case. This set-up gives more possibilities
for a modulation of the Casimir force which could be potentially
useful in micro- and nanosystems~\cite{Capasso:2007}. Besides,
dissimilar plates are required for getting a repulsive Casimir
force, the realization of which would constitute a major achievement
in the field. It is known~\cite{Kenneth:2002ij,Capasso:2003} that the force may change the sign if one
of the plates has nontrivial magnetic permeability, $\mu\ne1$. Novel
artificial materials~\cite{Pendry:1999,Veselago:2006} with magnetic
response arising from micro- or nanoinclusions have recently been
shown to become possible candidates for observing a repulsive
Casimir force\cite{Dalvit:2008}. Here we investigate the
contribution of the surface plasmon interaction between unequal
plates towards a better understanding of the nature of the Casimir
repulsion. The general conclusion of our study is that the surface
modes should be decoupled as much as possible in order to achieve
maximum repulsion.

The outline of the paper is the following. In Section II we
calculate the short and long distance asymptotes of the Casimir
force for nonequal mirrors. We show that if one of the mirrors has a
non-unity magnetic permeability the force is positive at short
distances provided the dielectric permittivity is non-unity as well.
It is known that at long and medium distances this set-up yields
repulsion~\cite{Henkel:2005,Pirozhenko:2008tr} if one of the mirrors
is more magnetic than dielectric. Modeling the dielectric
permittivity with a plasma model we find the minimal ratio between
magnetic and dielectric plasma frequencies required to get
repulsion. It is slightly larger than $1$. The case of a purely
dielectric mirror facing a purely magnetic one is known to yield the
largest possible repulsive force~\cite{Boyer:1974,Pirozhenko:2008tr}
and we consider it separately. There the force has an unusual short
distance asymptote, $F\sim1/L$, that is explained in Section III
where we study the plasmon contribution to the force between
nonequal mirrors. Section IV contains a discussion of the results
and conclusions.

\section{The Casimir force between dissimilar mirrors}

The Casimir force between two flat mirrors separated by a distance
$L$ is given by~\cite{Lifshitz:Eng:1956,Lambrecht:2000}
\begin{equation}
F(L)=-\frac{\hbar}{2\pi^2}\sum_{\rho}\int\limits_0^{\infty} d\kappa
\kappa^2 \int\limits_0^{c\kappa} d\omega \,\frac{r^{\rho}_A\, r^{\rho}_B}{\exp(2\kappa
L)-r^{\rho}_A\, r^{\rho}_B}.
\label{eq1}
\end{equation}
Here $r^{\rho}(i\omega,\kappa)$, $\rho=TE,TM$, are the reflection coefficients at
imaginary frequencies for the mirrors facing vacu\-um

\begin{equation}
r^{TM}_i(i\omega)=\frac{\kappa_i-\varepsilon_i \kappa}{\kappa_i+\varepsilon_i
\kappa},\quad r^{TE}_i(i\omega)=-\frac{\kappa_i-\mu_i \kappa}{\kappa_i+\mu_i
\kappa},
\label{eq2}
\end{equation}
with a dielectric permittivity
$\varepsilon_i=\varepsilon_i(i\omega)$, a magnetic permeability
$\mu_i=\mu_i(i\omega)$, the imaginary longitudinal wavevector
respectively in vacuum and in the material
$\kappa=\sqrt{\omega^2/c^2+k^2}$,
$\kappa_i=\sqrt{\omega^2/c^2\,(\varepsilon_i \mu_i-1)+\kappa^2}$ for
mirrors numbered by $i=A,B$.

The properties of the material enter the expression for the Casimir
force through the dielectric permittivity and magnetic permeability
at imaginary frequencies. From  the Kramers-Kronig causality
relation it follows~\cite{LandLif:EDCM:1984,Klimchitskaya:2007td},
that these functions are always real and positive
$\varepsilon(i\omega), \mu(i\omega)\geq1$. The sign of the force is
defined by the sign of the integrand in~(\ref{eq1}).  As $|r(i\omega,\kappa)|\leq
1$, a "mode" $\{\omega,\kappa\}$ gives a repulsive contribution to
the force if the corresponding reflection coefficients of the
mirrors $A$ and $B$ have opposite signs. This is only possible if
the mirrors are different, $r_A \neq r_B$, and  if at least one
mirror has a nontrivial magnetic permeability. The second condition
follows from the analysis of the reflection coefficients~(\ref{eq2})
at different frequencies.

The Casimir force between plates with a frequency dependent
reflectivity is usually calculated numerically by making use of the
available optical data or solid state physics models consistent with
the Kramers-Kronig relation~\cite{Lambrecht:2000}. Analytical
expressions for the Casimir force at short and long distances are
known for equal bulk mirrors with dielectric permittivity described
by a plasma, Drude or Lorenz models (see for,
example~\cite{Genet:2004}) and for mirrors of finite
thickness~\cite{pirozhenko:013811}. In what follows we derive
analytic expressions for the Casimir force at short and long
distances for dissimilar plates described by the models mentioned
above.

\subsection{Short distance asymptote}
The material properties define a characteristic length scale
$\lambda_\textrm{ch}$ for the scattering of the field on the plates
and thus also of the system. For example, if the mirrors optical
properties are described by a plasma model, this length scale
corresponds to the plasma wavelength. We therefore define the short
distance range with respect to this characteristic wave-length of
the permittivity model, $L\ll\lambda_\textrm{ch}$. As the main
contribution to the integral over $\kappa$~(\ref{eq1}) comes from
$\kappa\sim 1/L$, we deduce that only wave-vectors
$1/\kappa<<\lambda_\textrm{ch}$ contribute essentially to the
Casimir force. At short distances we may approximate the wave-vector
\begin{equation}
\kappa \sim k+\frac{1}{2}\frac{\rho_i^2}{k}+\dots \label{A2}
\end{equation}
where $\rho_i^2\equiv(\varepsilon_i\mu_i-1)\omega^2/c^2$, $i=A,B$.
The force is then given by
\begin{eqnarray}
F=-\sum_{\rho}\frac{\hbar}{2\pi^2}\int\limits_0^{\infty}dk \,k^2
\int\limits_0^{\infty} d\omega \frac{r_A^{\rho} r_B^{\rho}\, e^{-2 k
L}}{1-r_A^{\rho} r_B^{\rho}\, e^{-2 k L}} \label{A1}\\
r^{TM}_i(i\omega)=\frac{1-\varepsilon_i}{1+\varepsilon_i}+\frac{\varepsilon_i}{(1+\varepsilon_i)^2}
\frac{\rho_i^2}{k^2}+\dots,\nonumber\\
r^{TE}_i(i\omega)=\frac{1-\mu_i}{1+\mu_i}+\frac{\mu_i}{(1+\mu_i)^2}
\frac{\rho_i^2}{k^2}+\dots. \label{A3}
\end{eqnarray}
Thus the magnetic/dielectric properties of the material show up in
the second therm of the expansion of the transverse
magnetic/electric reflection coefficient~(\ref{A3}). This term is
usually omitted. However, here we will take it into account in order
to determine the sign of the force when $\varepsilon$ or $\mu$ are
equal to unity and the leading term vanishes.

Let the material be described by the general model
\begin{eqnarray}
\varepsilon(\omega)=1-\frac{\omega_{e}^2}{\omega^2-\omega_0^2+i\Omega_{e}\omega}\equiv
1+\frac{\omega_e^2}{\omega^2}\, B_e(\omega)\nonumber\\
\mu(\omega)=1-\frac{\omega_{m}^2}{\omega^2-\omega_0^2+i\Omega_{m}\omega}\equiv
1+\frac{\omega_m^2}{\omega^2}\, B_m(\omega), \label{A4}
\end{eqnarray}
which reduces to the well known plasma model for
$\Omega_e=\omega_0=0$ with a plasma frequency $\omega_e$ and to the
Drude model with $\omega_0=0$ where $\Omega_e$ describes the
electronic relaxation frequency. Then the leading terms of the short
distance asymptote for the reflection coefficients are given by
\begin{eqnarray}
r^{TM}(i\omega)&=&-\frac{\omega_{e}^2}{2(\omega^2+\omega_0^2+\Omega_{e}\omega)+\omega_{e}^2},\nonumber \\
r^{TE}(i\omega)&=&\frac{\omega_{m}^2}{2(\omega^2+\omega_0^2+\Omega_{m}\omega)+\omega_{m}^2}.
\label{A6}
\end{eqnarray}
The absorbtion in the material has the strongest influence at low
frequencies. On the contrary, in the short distance range mainly the
high frequencies make the decisive contribution to the force.
Therefore we may neglect the absorbtion in the material for short
distances, by putting $\Omega_e=\Omega_m=0$.

First we calculate the short distance limit for the force between a
purely magnetic mirror A facing a purely dielectric mirror B. When
$\varepsilon=1$, the leading term in $r^{TM}$ vanishes,
$\rho^2=\omega_m^2 B_m(i\omega)/c^2$, and
\begin{equation}
r^{TM}(i\omega)=\frac{\omega_m^2}{4\,c^2}\,
\frac{\omega^2}{k^2}\,\frac{1}{\omega^2+\omega_0^2+\Omega_m\omega}.
\label{A7}
\end{equation}
To assess the TM contribution, we therefore use the reflection
coefficients $r^{TM}_A$ defined by (\ref{A6}) and  $r^{TM}_B$
defined by (\ref{A7}).

In the same manner, when  $\mu=1$, then the leading term in $r^{TE}$
vanishes, $\rho^2=\omega_e^2 B_e(i\omega)/c^2$ and
\begin{equation}
r^{TE}(i\omega)=-\frac{\omega_e^2}{4\,c^2}\,
\frac{\omega^2}{k^2}\,\frac{1}{\omega^2+\omega_0^2+\Omega_e\omega}.
\label{A8}
\end{equation}
For the contribution coming from the TE modes, the reflection
coefficients $r^{TE}_B$ given by expression (\ref{A8}), and
$r^{TE}_A$ defined by eqn.~(\ref{A6}) must be employed.

We then evaluate the force for the Drude model, i.e. $\omega_0=0$.
The TM contribution  reads
\begin{equation}
F^{TM}=\frac{\hbar \sqrt{2}\,\omega^2_{mA}\,\omega_{eB}}{32 \pi c^2\,L} \int_0^{\infty} dk
\frac{k\,e^{-2 k }}{\sqrt{k^2+\frac{\omega^2_{mA} L^2}{8 c^2}\,e^{-2 k}}}.
\label{A10}
\end{equation}
where we can neglect the second term in the denominator, as
$\omega^2_{mA} L^2/4c^2<<1$. The TM contribution to the Casimir
force then simplifies to
\begin{equation}
F^{TM}\approx\frac{ \sqrt{2}}{64}\, \frac{\hbar}{\pi c^2} \,\frac{\omega^2_{mA} \omega_{eB}}{L}.
\label{A11}
\end{equation}
The contribution of TE modes is calculated in an analogous way,
replacing $\omega_{mA}$ in (\ref{A11}) by $\omega_{eB}$. This leads
to the following expression for the total Casimir force
\begin{equation}
F\approx\frac{ \sqrt{2}}{64}\, \frac{\hbar}{\pi c^2}\,\frac{(\omega^2_{eB}\,\omega_{mA}+\omega_{mA}^2 \,\omega_{eB})}{L}.
\label{A12}
\end{equation}
The force is repulsive and has an unusual short distance asymptote
$\sim 1/L$. This unusual behavior will be explained in the next
section.

%________________________________________________

Now we evaluate how the introduction of a small dielectric
permittivity for the former purely magnetic mirror B affects the
Casimir force. To this aim we calculate the short distance limit for
the force between a magneto-dielectric mirror A facing a purely
dielectric mirror B. In this case the TE contribution at short
distances is negligible with respect to the TM one. The TM
contribution to the Casimir force has now to be evaluated using the
TM reflection coefficients~(\ref{A6}). After
expanding the integrand in~(\ref{A1}) and integration over $k$ we
obtain $F=-H_{AB}/3L^3$, where $H_{AB}$ is sometimes refereed in the
literature as non-retarded Hamaker constant~\cite{Bergstrom:1997}. At zero temperature it is
\begin{eqnarray}
H_{AB}&=&\frac{3\hbar}{8\pi^2}
\int\limits_0^{\infty} d\omega\sum_{n=1}^{\infty}\frac{(r_A^{TM}[i\omega]\,r_B^{TM}[i\omega])^n}{n^3}.
\label{A13}
\end{eqnarray}
Further we recast~(\ref{A13}) in order to see the dependence of force on the parameters more clearly.
If we rewrite the reflection coefficients~(\ref{A6}) as
$$r^{TM}_i(i\omega)=-\frac{\Omega_{1,i}^2}{\omega^2+\Omega_{2,i}^2}=-\frac{\Omega_{1,i}^2}{\Omega_{2,i}^2}
\frac{\Omega_{2,i}^2}{\omega^2+\Omega_{2,i}^2},\quad i=A,B,$$
where $\Omega_{1,i}^2=\omega_{e,i}^2/2$, and
 $\Omega_{2,i}^2=\omega_{e,i}^2/2+\omega_0^2$, the integrand
simplifies. The result of integration over $\omega$ may be expressed
in terms of the hypergeometric series.

Finally we obtain the  Casimir force at short distances
\begin{eqnarray}
F&\simeq& -\frac{\hbar}{L^3}
\frac{\Omega_{2B}}{16 \pi^2}\sum_{k=0}^{\infty}\frac{\Gamma(k+\frac{1}{2})}{k!}\,G_k\,\left(1-
\frac{\Omega_{2B}^2}{\Omega_{2A}^2}\right)^k, \label{A18}
\end{eqnarray}
where
\begin{eqnarray}
G_k&=&\sum_{n=1}^{\infty}\frac{\Gamma(n+k)\,\Gamma(2n-\frac{1}{2})}{\Gamma(n)\;\Gamma(2n+k)\,n^3}
\left(\frac{\Omega_{1A}}{\Omega_{2A}}
\frac{\Omega_{1B}}{\Omega_{2B}} \right)^{2n}.
\label{A18a}
\end{eqnarray}

If the plates are described by a plasma model we have
$\omega_{0A}=\omega_{0B}=0$, $\Omega_{1A}={\Omega_{2A}}$,
$\Omega_{1B}={\Omega_{2B}}$ while  the numerical coefficients
(\ref{A18a}) do not depend on the plasma frequencies of the mirrors.
The Casimir force then writes
$$F\simeq -\frac{\hbar}{16 \pi^2 L^3}
\frac{\omega_{e,B}}{\sqrt{2}}\;\gamma,$$ where, for example,
$\gamma\approx1.744$ for two equal mirrors. If, in contrast, the
plasma frequency of mirror A is slightly higher,
$\omega_{e,B}/\omega_{e,A}=0.9$, the force increases, and
$\gamma\approx1.836$. The result does not depend on the magnetic
properties of the mirror A.

Our analysis has shown that the Casimir force between a mirror A
with $\mu_B\ne1$ and a purely dielectric mirror B is always
attractive in the short distance range and determined by the modes
corresponding to the TM polarization of the electromagnetic field.
When $\varepsilon_A\to 1$ the TM contribution becomes small and
comparable to the contribution of the TE modes. Only in this case
the force is repulsive at short distances and has the form defined
in~Eq.(\ref{A12})

\subsection{Long distance asymptote}
In order to find the condition for the occurrence of a repulsive
Casimir force we add here the analysis of the long distance limit
between the two mirrors ($L>>\lambda_{eA},\lambda_{eB}$). For two
dissimilar purely dielectric mirrors described by a plasma model the
long distance limit is obtained by expanding the integrand
of~(\ref{eq1}) in powers of the small parameter $\lambda_{eA}/L$ or
$\lambda_{eB}/L$
\begin{equation}
F|_{L>>\lambda_{eA},\,\lambda_{eB}}=\eta \,F_{C}(L), \quad
\eta\approx1-\frac{4}{3\pi}\frac{\lambda_{eA}+\lambda_{eB}}{L}.
\end{equation}
where we have introduced the force reduction factor
$\eta$\cite{Lambrecht:2000}. For a magneto-dielectric mirror A
described by a plasma model with plasma frequencies
$\omega_{mA},\omega_{eA}$ in front of a purely dielectric mirror B,
characterized by a plasma frequency $\omega_{eB}$, the force
reduction factor tends at long distances to
\begin{eqnarray}
\eta(\alpha)\to \frac{180}{\pi^4}\frac{3}{8}\left\{\alpha\int\limits_0^{1/\alpha}
d\Omega \,\mbox{Li}_4\left(\frac{\Omega-1}{\Omega+1}\right)\right.\nonumber\\
+\left.\frac{1}{\alpha}\int\limits_0^{\alpha}d\Omega\sum_{n=1}^{\infty}
\frac{(-1)^n}{n^4}\left(\frac{\Omega-1}{\Omega+1}\right)^n \right\}
\label{long_dist}
\end{eqnarray}
$\alpha$ gives the ration between the magnetic and dielectric plasma
frequency of mirror A $\alpha=\omega_{mA}/\omega_{eA}$. In contrast
to the case of two dielectric mirrors, the long distance limit of
the reduction factor  between a magneto-dielectric and a pure
dielectric mirror is not $1$. In particular the sign of the Casimir
force is defined by the parameter $\alpha$. As the leading
term~(\ref{long_dist}) of the long distance asymptote does not
depend on the plasma frequency of the nonmagnetic mirror B, we can
estimate the ratio between the dielectric and magnetic plasma
frequencies of the plate A, required to get a repulsive force. A
numerical analysis shows that for a value of $\alpha_0\approx1.0255$
the long distance asymptote of the Casimir force vanishes. In other
words, within the plasma model the Casimir repulsion is achieved
when $\alpha>\alpha_0=1.0255$. The ratio is lower than the one obtained
in~\cite{Kenneth:2002ij} for \textit{constant} dielectric permittivity
and magnetic permeability, $\mu \sim 1.08\varepsilon$, provided mirror B is a perfect conductor.

On the other hand, when mirror A has equal dielectric and magnetic
responses, that is $\alpha=1$, the force is positive at any plate
separation and the long distance limit of the
 reduction factor is given by
 $$
 \eta(1)=\frac{180}{\pi^4}\frac{3}{64}\int\limits_0^{1}
d\Omega \,\mbox{Li}_4\left(\left(\frac{\Omega-1}{\Omega+1}\right)^2\right)=0.0205
 $$
Finally, for a purely dielectric mirror A facing a purely magnetic
mirror B we obtain the long distance limit characterized by strong
repulsion
\begin{equation}
F|_{L>>\lambda_{eA},\,\lambda_{eB}}=\eta \,F_{Cas}(L), \quad
\eta\approx-\frac{7}{8}+\frac{7}{6\pi}\frac{\lambda_{eA}+\lambda_{mB}}{L}.
\end{equation}
The repulsive force given by the first term of this expansion was
first obtained by Boyer~\cite{Boyer:1974} for two non-dispersive
mirrors with $\epsilon_A=\infty$, $\mu_A=1$ and $\varepsilon_B=1$,
$\mu_B=\infty$.

\section{Description in terms of interacting surface plasmons}

In this section we will analyze the attractive and repulsive
behavior of the Casimir force taking into account the differences in
the surface plasmon coupling for the different combinations of
materials. Eq. (\ref{eq1}) for the Casimir force includes both
photon and plasmon contributions that have different signs.
In~\cite{IntraLam:2005} it was shown that the value of the Casimir
energy is the result of the compensation between the photon and
plasmon contributions for equal mirrors described by plasma model.
Moreover at short distances the force is entirely defined by the
attraction of the surface plasmons. The analysis of the photon and
plasmon contributions for dissimilar mirrors, especially if they
have a nontrivial magnetic permeability, is more complicated. That
is why we restrict ourselves to the plasma model for the material
properties of the mirrors. We start from the general definition of
surface plasmon energy. As in the previous section, we first
consider the plasmon interaction for two dissimilar purely
dielectric mirrors, then for magneto-dielectric mirrors and finally
we discuss a purely dielectric mirror facing a purely magnetic one.

Between two arbitrary plates the surface plasmon modes exist for
both field polarizations, TE and TM. Their frequencies
$\omega^{\rho}_{\sigma}$ are implicitly defined as the solutions of
the equations
\begin{equation}
\prod\limits_{i=A,B}^{}\frac{\kappa_i+\varepsilon_i q}{\kappa_i-\varepsilon_i q}=e^{-2\,q\,L},\quad \prod\limits_{i=A,B}^{}\frac{\kappa_i+\mu_i q}{\kappa_i-\mu_i q}=e^{-2\,q\,L}
\label{eq_nonid}
\end{equation}
with $ q^2=k^2-\omega^2/c^2\geq0,\;
\kappa^2_{i}=k^2-\varepsilon_{i}\mu_{i}\,\omega^2/c^2, \quad i=A,B$.
The left and right equations refer respectively to the TM and TE
polarization. Both equations have two different solutions which are
numbered by the index $\sigma$.

The vacuum energy of the interacting surface plasmons living on the
plane mirrors is then formally given by
\begin{equation}
E^{sp}=\frac{\hbar}{2}\sum_{\rho,\,\sigma}^{}\int\limits_{k_{(\sigma)}}^{\infty}\frac{d
k\, k}{2\pi} \left[\omega^{\rho}_{\sigma}\right]^{L}_{L\to\infty},
\quad \rho=TE, TM. \label{energy_plasm}
\end{equation}
The infinite energy of single-surface plasmons has been subtracted
in this expression as it corresponds to an infinite plate
separation. When both mirrors are equal, the two TM surface plasmons
which one obtains as solutions of (\ref{eq_nonid}) are called
symmetric and antisymmetric
plasmons~\cite{IntraLam:2005,Bordag:2005} or binding and
anti-binding resonances~\cite{HenGreffet:2004}. Each solution exists
for $k>k_{(\sigma)}$.

\subsection{Two dissimilar purely dielectric mirrors}
\begin{figure}[t]
\epsfig{file=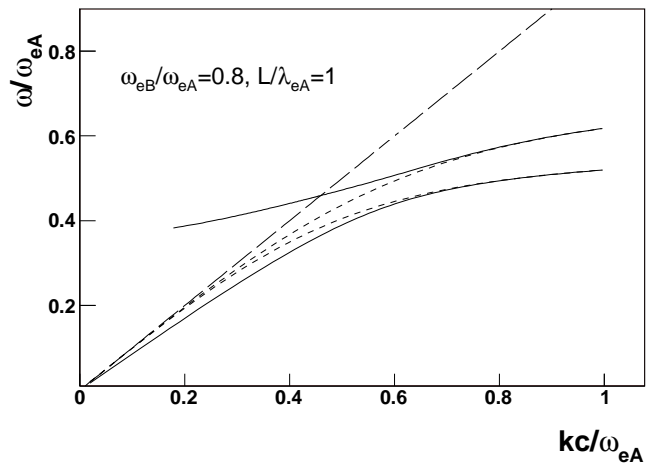,width=9.5cm} \caption{Solid curves:  the
TM plasmon modes which are the solution of~(\ref{eq_nonid}); dashed
curves: noninteracting single-surface plasmons, (\ref{plasm_B});
long-dashed line: boundary between propagative and evanescent
sectors} \label{fig2}
\end{figure}

Let us first consider two dissimilar dielectric mirrors described by
a plasma model
\begin{eqnarray*}
\varepsilon_i(\omega)=1-\frac{\omega_{e,i}^2}{\omega^2}, && \mu_i(\omega)=1,\; i=A,B
\end{eqnarray*}
with $\omega_{eB}/\omega_{eA}=\beta_e$. In TM polarization on each mirror lives a
single-surface plasmon with a frequency
\begin{equation}
\omega^{sp}_{i}=\frac{1}{\sqrt{2}}\left[\omega_{ei}^2+2|k|^2
c^2-\sqrt{\omega_{ei}^4+4 k^4 c^4}\right]^{1/2}  \label{plasm_B}
\end{equation}
When both mirrors become close to each other they start to interact according to TM
equation (\ref{eq_nonid}). It has two solutions.
In analogy with the case of two equal mirrors we call them symmetric and antisymmetric
plasmons. Let $\omega_{eB}\leq\omega_{eA}$ ($\beta_e\leq1$). At large plate
separations the frequency of the symmetric
plasmon, $\omega_{+}$, tends to the frequency of the unperturbed
surface plasmon of mirror A $\omega^{sp}_{A}$, while the frequency of
the antisymmetric one, $\omega_{-}$, approaches the frequency of the
unperturbed plasmon of mirror B, $\omega^{sp}_{B}$. The behavior of
the plasmon modes being the solutions of~(\ref{eq_nonid}) is plotted
on Fig.~\ref{fig2} for  $L/\lambda_{eA}=1$ and $\beta_e=0.8$.
The straight line $\omega=kc$  marks the limit between the sector of
propagating waves $kc<\omega$ and evanescent waves $kc>\omega$. The
plasmon $\omega_{+}$ crosses the boundary between the propagative
and evanescent sectors when
\begin{equation}
k\equiv
k_{(+)}=\frac{\omega_{eA}}{c}\sqrt{\frac{\beta_e(\beta_e+1)}{1+\beta_e(\Lambda+1)}},
\quad \Lambda=\omega_{eA}L/c. \label{tail}
\end{equation}
The plasmon mode $\omega_{-}$ lies entirely in the evanescent
sector, $k_{(-)}=0$.

In~\cite{IntraLam:2005} and the subsequent papers
\cite{IntraLam:2006,Intravaia:2007} an adiabatic mode definition was
used which attributes the entire mode $\omega_{+}$ to the evanescent
sector. Mathematically it is equivalent to putting $k_{(+)}=0$. This
choice gives a repulsive plasmon contribution to the Casimir force
while the photon contribution is attractive, and allows to recover
the perfect mirrors limit with an attractive force coming alone from
the photonic mode contribution corresponding to the original
derivation by H. Casimir\cite{Casimir:1948}. For simplicity we will
adopt here the  convention of \cite{Bordag:2005,Lenac:2006} and
attribute the two parts of the symmetric surface plasmon to the
corresponding sectors. When calculating its vacuum energy we start
the integration over $k$ from the value given in~(\ref{tail}).

\begin{figure}[btp]
\epsfig{file=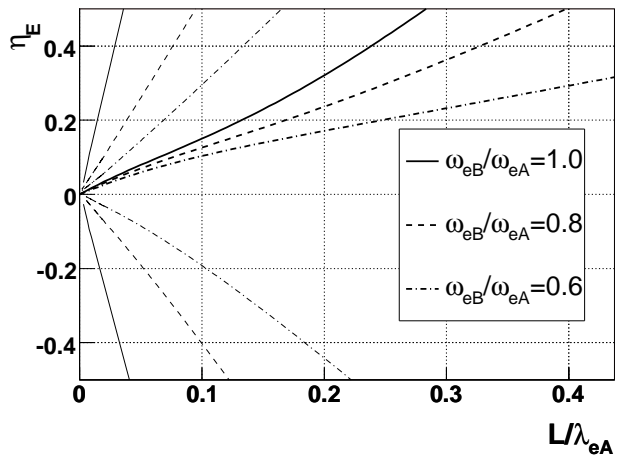,width=9.0cm} \caption{Thick
solid, dashed and dash-dotted lines: the normalized contribution of
the surface plasmons to Casimir energy for different ratios
$\omega_{eB}/\omega_{eA}$. The attractive contribution of
$\omega_{-}$ and repulsive contribution of $\omega_{+}$ to
$\eta_{pl}=E^{sp}{}/E_C$ are traced in thin lines.} \label{fig3}
\end{figure}

Fig.~\ref{fig3} shows a plot of the surface plasmon contribution to
the energy for different values of $\beta$. The total plasmon energy
is negative corresponding to an attractive force (positive $\eta_{pl}$)
and comprises the repulsive contribution of the symmetric plasmon
and the attractive contribution of the antisymmetric plasmon. The
attraction between the surface plasmons is the strongest when the
plasma frequencies of the plates coincide.

\begin{figure}[btp]
\epsfig{file=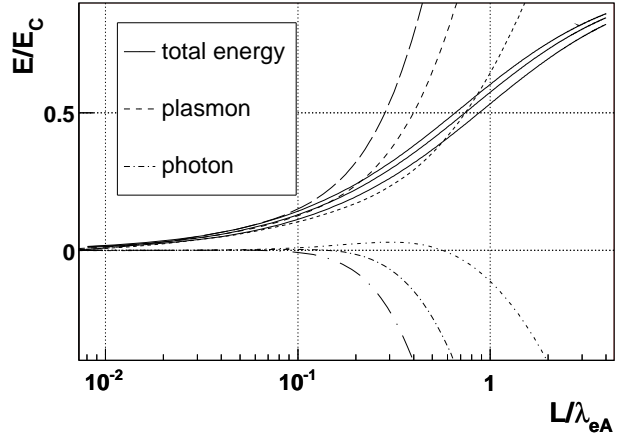,width=9.0cm}
\caption{The Casimir energy for the dissimilar mirrors as a result of
cancelation between photon and plasmon contributions.
The plasmon contributions  for
$\omega_{eB}/\omega_{eA}=1,0.8,0.6$ are plotted respectively by long, medium, and short dash lines.
The dash-dotted curves give the corresponding photon contribution, and the solid lines give the
total Casimir energy. The total energy is the largest for equal mirrors, $\alpha=1$,
the lowest plot is the total energy for $\alpha=0.6$.}
\label{fig4}
\end{figure}

Fig.~\ref{fig4} shows plasmon and photon contributions to the
Casimir energy as a function of the normalized distance
$L/\lambda_{eA}$ between the plates for three different ratios of
the electric plasma frequencies, namely $\omega_{eB}/\omega_{eA}=1$
(long dashed and dotted-dashed), $\omega_{eB}/\omega_{eA}=0.8$
(medium dashed and dotted dashed) and $\omega_{eB}/\omega_{eA}=0.6$
(short dashed and dotted dashed). When the plates are unequal,
$\omega_{eA}\ne\omega_{eB}$, the balance between the plasmon and
photon contribution to the energy obviously changes. Each
contribution on its own decreases and so does the total Casimir
energy. As the plasma frequencies do not match, the plasmon
interaction is weaker and the respective contribution to the energy
is smaller.

On the other hand, the photon contribution is reduced as well. The
transparency of the mirror for the propagating waves is governed by
its plasma frequency. If $\omega_{eA}>\omega_{eB}$, the fluctuations
from the band $\omega_{eB}<\omega<\omega_{eA}$ are reflected by the
plate A, but pass through the plate B. Thus the high frequency
cut-off for the cavity formed by both mirrors is defined by the
plasma frequency of the less reflecting plate B. Consequently the
distance depending part of the vacuum energy which comes from the
propagative sector has a smaller value than in the case of plates
with equal plasma frequencies $\omega_{eA}$, resulting in a smaller
Casimir force.

The long distance asymptote of the plasmon contribution,
$\Omega_{e,i}\equiv L\omega_{e,i}/c\gg1$ may be calculated
explicitly. To this end, we first introduce the dimensionless
variables $K=kL, \Omega=\omega L/c$, and
$\Omega_{e,i}=\omega_{e,i}L/c$ in
(\ref{energy_plasm},{\ref{eq_nonid})  leading to
\begin{equation}
E^{sp}=\frac{\hbar c}{4 \pi L^3}\left\{\int\limits_{K_{+}}^{\infty} d
K K (\Omega_{+}-\Omega_{A}^{sp})+\int\limits_{0}^{\infty} dK K(\Omega_{-}-\Omega_{B}^{sp})\right\}.
\label{energy_plasm_dimless}
\end{equation}
Then we change the integration variable in
(\ref{energy_plasm_dimless}), $K\rightarrow Q=\sqrt{K^2-\Omega^2}$
and write the renormalized energies of the symmetric and
antisymmetric plasmons as
\begin{eqnarray}
\bigl.E^{sp}\bigr|_{\Omega_{eB}\gg1}&=&\frac{\hbar c}{4 \pi L^3}\left\{\int\limits_0^{\infty} dQ Q \{\Omega_{-}+\Omega_{+}-\Omega_{B}^{sp}-\Omega_{A}^{sp}\},\right. \nonumber\\
&+&\left.\int\limits_{0}^{Q_A}dQ \,Q\, \Omega_A^{sp}
+\frac{1}{3}(\Omega_{+}^3-\Omega_A^3)\bigl|_{K_{(+)}}^{\infty}\bigr.\right\}.\label{sp_long}
\end{eqnarray}
If the plasma frequencies of the plates are close to each other, the
following simplifications can be performed
\begin{eqnarray*}
&&\Omega_{\pm}=\sqrt{\frac{\Omega_{eB}\,Q}{2\beta}}\sqrt{\frac{1+e^{-2Q}}{1-e^{-2Q}}} \, f_{\pm}^{1/2}(\beta_e,Q),\quad 0<\beta\leq1, \\
&&f_{\pm}(\beta,Q)=(1+\beta_e)\pm\sqrt{(1-\beta)^2+\frac{16\beta e^{-2Q}}{(1+e^{-2Q})^2}},\\
&&\Omega_{A}^{sp}=\sqrt{\beta^{-1}\Omega_{eB}Q},\quad \Omega_{B}^{sp}=\sqrt{\Omega_{eB}Q}, \\ &&K_{(+)}=\sqrt{1+\beta^{-1}}\sqrt{\Omega_{pB}},\\
&& Q_A=\sqrt{K_{(+)}^2-\Omega_{A}^{sp}(K_{(+)})}\to1+\beta,
\end{eqnarray*}
With these we find a long distance approximation for the  vacuum
energy of the interacting surface plasmons
\begin{equation}
\bigl.E_{sp}\bigr|_{\Omega_{eB}\gg1}=
\frac{\hbar c\sqrt{\Omega_{eB}}}{4 \sqrt{\beta_e}\pi L^3} \psi(\beta)=
\frac{\hbar \sqrt{c\omega_{eA}}}{4 \pi L^{5/2}} \psi(\beta_e).
\label{plasm_long_approx}
\end{equation}
For the different values of $\beta_e$ plotted in Fig.\ref{fig4} we find
$\psi(0.6)\approx-0.0663$, $\psi(0.8)\approx-0.163$ and
$\psi(1)=-0.2798$. The last value, corresponding to equal mirrors,
confirms the result published in \cite{Bordag:2005}. It is important
to note that the accuracy of the long distance approximation
~(\ref{plasm_long_approx}) becomes worse as $\beta$ decreases.

The situation when one of the plates is perfectly conducting
corresponds to $\beta_e\ll1$ (but $\Omega_{eB},\Omega_{eA}>>1$). In
this case  the anti-binding surface plasmon $\omega_{+}$  dominates
at large distances. The total energy of the surface plasmons becomes
positive, yielding repulsion.

\subsection{Two dissimilar magneto-dielectric mirrors}
To assess under which conditions Casimir repulsion may arise we will
finally investigate the situation of two dissimilar
magneto-dielectric plates.  Let the dielectric permittivity and
magnetic permeability of the mirrors be described by a plasma model
\begin{eqnarray*}
\varepsilon_i(\omega)=1-\frac{\omega_{e,i}^2}{\omega^2}, && \mu_i(\omega)=1-\frac{\omega_{m,i}^2}{\omega^2},\; i=A,B
\end{eqnarray*}
Then the frequencies of the single surface plasmons of the
respective mirror are given by
\begin{equation}
\omega^{TM}_{sp,i}=\frac{\omega_{e,i}}{\sqrt{2}}\left[1-\frac{2 k^2 c^2}{\omega_{m}^2-\omega_{e}^2}-
\sqrt{1+\frac{4 k^4 c^4}{(\omega_{m}^2-\omega_{e}^2)^2}}\right]^{1/2}_i.
\label{plasm_nonid}
\end{equation}
When $k\to 0$ the single surface plasmon frequency vanishes, while
it tends to $\omega_{e,i}/\sqrt{2}$ when $k\to\infty$. If
$\omega_{m}=\omega_{e}\equiv\omega_{p}$, the frequency of the single
surface plasmon does not depend on $k$ and
$\omega^{TM}_{sp,i}\to\omega_{p,i}/\sqrt{2}$. In the limit of small
plate separations, corresponding to large wave vectors, the
equations for the interacting surface plasmons may be solved
explicitly:
\begin{equation}
(\omega^{TM}_{\pm})^2=\frac{\omega_{eA}^2+\omega_{eB}^2}{4}\mp
\sqrt{\frac{(\omega_{eA}^2-\omega_{eB}^2)^2}{16}+\frac{\omega_{eA}^2
\omega_{eB}^2}{4} e^{-2 |k| L}} \label{plasm_nonid_short}
\end{equation}
To obtain the solutions corresponding to the TE surface plasmons one
has to replace $\omega_{e,i}$ by $\omega_{m,i}$ for both mirrors
$i=A,B$.

Substituting (\ref{plasm_nonid_short}) into (\ref{energy_plasm}) we
may derive the vacuum energy of the surface plasmons at short plate
separation. It is given by
\begin{eqnarray}
E_{pl}(L)=\frac{\hbar}{16\pi
L^2}\left[\frac{\omega_{eA}}{\sqrt{2}}\chi(\beta_e)+
\frac{\omega_{mA}}{\sqrt{2}}\chi(\beta_m)\right], \label{pl_asymp}
\end{eqnarray}
where the parameters $\beta_e=\omega_{eB}/\omega_{eA}$ and
$\beta_m=\omega_{mB}/\omega_{mA}$ are introduced. $\chi(z)$ is given
by the following expression
\begin{eqnarray}
\chi(z)=\int\limits_0^{\infty}dk k\left\{\left[\left(z_{+}^2+
\sqrt{z_{-}^4+z^2\,e^{-k}}\right)^{1/2}-1\right]\nonumber\right.\\+
\left.\left[\left(z_{+}^2-
\sqrt{z_{-}^4+z^2\,e^{-k}}\right)^{1/2}-z\right]
\right\}
\end{eqnarray}
with $z_{+}^2=(1+z^2)/2$ and $z_{-}^2=(1-z^2)/2$. For a positive
argument the function $\chi$ is negative and varies from $0$ to
$\chi(\infty)\to-0.1358$. Thus the energy of interacting surface
plasmons is always negative yielding attraction between unequal
magneto-dielectric mirrors. Reformulating the argument in terms of
the Casimir force, $F_{pl}=-dE_{pl}/dL$, it means that
magneto-dielectric plates allow only for an attractive Casimir force
at short distances, which is the result of the interaction of the
surface plasmons.

The limit $z\to\infty$ corresponds to vanishing frequencies
$\omega_{eA}$ or $\omega_{mA}$ and hence to vanishing TE and TM
plasmon contributions to the Casimir energy. For equal mirrors we
find $z=1$ and $\chi(1)\simeq-0.2776$, corresponding to the maximum
value of the Casimir energy.

\begin{figure}[btp]
\epsfig{file=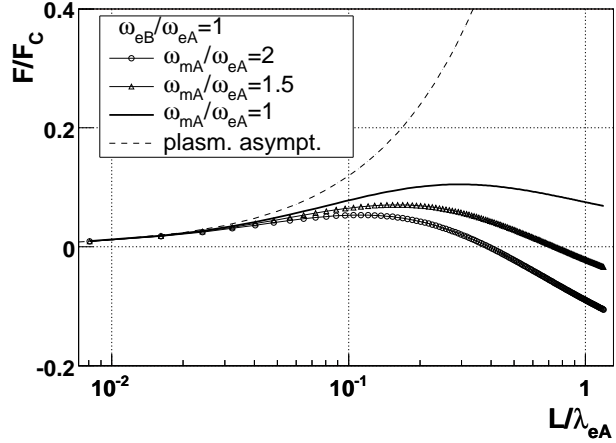,width=9.0cm}
\caption{Reduction factor $\eta_F=F/F_C$ at short plate separations as a function of dimensionless
distance $\Lambda=2\pi L/\lambda_{eA}$. The dashed line is the short distance asymptote of the surface
plasmon interaction.}
\label{fig1}
\end{figure}

In Fig.\ref{fig1} we finally show the result for the force
correction factor $\eta_F(L/\lambda_{eA})=F/F_C$, evaluated
numerically as a function of the distance normalized by
$\lambda_{eA}=2\pi c/\omega_{eA}$. The force is calculated between
two plates with
$\varepsilon_A(\omega)=\varepsilon_B(\omega)=1-\omega_{eA}^2/\omega^2$.
Plate B is purely dielectric. The magnetic permeability of plate A
is $\mu_A(\omega)=1-\omega_{mA}^2/\omega^2$. We vary the magnetic
response of the plate A and calculate the force making use of the
exact formula~(\ref{A1}). The total correction factor $\eta_F$ is
strongly affected by $\omega_{mA}$ and becomes negative at medium
and long distances if $\omega_{mA}>\omega_{eA}$, which indicates a
repulsive Casimir force. The TM plasmons exist on both plates, while
the TE plasmons live only on plate A. Thus their energy does not
depend on the plate separation and they do not influence the force.

The plot also shows the short distance asymptote of the plasmon
contribution (dashed line), which is linear in
$\Lambda=L/\lambda_{eA}$, $\eta_F^{pl}\simeq 1.1933 \,\Lambda$ in
agreement with \cite{Lambrecht:2000}. It does not depend on the
magnetic permeability of plate A.

Finally we discuss the particular case of a purely dielectric mirror
facing purely magnetic one. From~(\ref{plasm_nonid}) one can easily
see that the mirror has no TM plasmon modes if it is purely magnetic
($\omega^2_{eA}=0$). Similarly if the mirror is purely dielectric
($\omega^2_{mB}=0$) it has no TE plasmon modes. Imagine a hypothetic
Casimir set-up with a purely dielectric mirror placed in front of a
purely magnetic one. Then the existing TM-plasmon mode of purely
dielectric mirror and TE-pasmon mode of the purely magnetic mirror
are not coupled.

Indeed, in this particular case the solution
(\ref{plasm_nonid_short}) does not depend on the distance between
the plates , $\omega^{TM}_{+}=0$,
$\omega^{TM}_{-}=\omega_{eB}/\sqrt{2}$ and
$\omega^{TE}_{+}=\omega_{mA}/\sqrt{2}$, $\omega^{TE}_{-}=0$. The
vacuum  energy is distance independent as well. Therefore the
plasmon modes do not contribute to the Casimir force. The force is
repulsive and entirely defined by the propagating modes. This
explains the unusual short distance asymptote~(\ref{A18}).

We have studied the interaction of the surface modes for the magneto-dielectric
plates described by the simplest plasma model. Surface polaritons living on the \textit{single}
interface of two media, one of which is left-handed with the magnetic permeability described by
alternative model,  was examined in~\cite{Ruppin:2000}. The interaction of these modes within the
Casimir cavity is of interest.

\section{Discussion and conclusions}
In the present paper we have considered two dissimilar plates with
frequency dependent dielectric permittivity and magnetic
permeability. Analytic expressions for the force at short and long
distances were derived. They give us the the sign of the force and
its dependence on the parameters  of the solid state physics models
describing the material of the plates. To explain these results we
have studied the vacuum energy of the interacting surface modes
living on the mirrors. We confirm the conclusion formulated
in~\cite{IntraLam:2005,Bordag:2005} for equal mirrors described by
plasma model,  that at all distances the value of the force is the
result of large cancelation between the plasmon and photon
contributions. The problem for dissimilar mirrors  appeared to be
more complicated. Varying the properties of plates affects both
propagative and evanescent modes. However our analysis indicates
that provided one of the mirrors is mainly magnetic the balance
between the plasmon and photon contributions is shifted to repulsion
and we have determined the ratio between the dielectric and magnetic
plasma frequencies required to obtain repulsion.  The general
conclusion of our study is that in order to achieve a repulsive
Casimir force, the surface plasmons should be decoupled as much
possible. The limiting, though experimentally not achievable, case
corresponds to a purely dielectric mirror facing a purely magnetic
one. For this situation we explain the repulsion at all distances by
the complete absence of the interaction between the surface modes.

\section{Acknowledgements}
We acknowledge financial support from the European Contract No. STRP 12142 NANOCASE.
I.P. is grateful for partial financial support from RFBR Grant No. 06-01-00120-\verb"a".

\end{document}